\documentstyle[preprint,aps]{revtex}

\begin{document}
\preprint{UNDPDK-97-01}
\title{Bounds on Dark Matter from the ``Atmospheric Neutrino Anomaly''
}
\author{J.M.~LoSecco}
\address{University of Notre Dame, Notre Dame, Indiana 46556}
\date{April 15 1997}
\maketitle
\widetext
\begin{abstract}
Bounds are derived on the cross section, flux and energy density of
new particles that may be responsible for the atmospheric
neutrino anomaly.\\
$4.6 \times 10^{-45}$ cm$^{2}$ $< \sigma <2.4 \times 10^{-34}$ cm$^{2}$\\
Decay of primordial homogeneous dark matter can be excluded.
\\
Subject headings: Cosmology --- Elementary Particles --- Dark Matter\\
\end{abstract}


\section{Introduction}
The atmospheric neutrino anomaly\cite{haines,kamioka,imbo} refers to
indications that the ratio
of muon neutrinos to electron neutrinos observed in underground detectors
is significantly less than expected.  Expectations are based on calculations
of neutrino fluxes derived from pions produced by primary cosmic ray
interactions in the upper atmosphere.  Since the earth's atmosphere is
not particularly dense the pions decay to produce muons and neutrinos
and a good fraction of the muons also decay to produce neutrinos.
Estimates based on detailed production models put the ratio of muon
neutrino flux to electron neutrino flux close to 2\cite{flux}.  The overall
normalization of these models is uncertain and so, it is the ratio for which
one has the highest confidence.

The most popular explanation for the deficiency is that neutrino oscillations
have converted some of the muon neutrinos to some other type so that the
muon neutrino flux observed is much closer to the electron neutrino flux.
The neutrino oscillation hypothesis has failed to be confirmed by a number of
experiments\cite{iupw,iclark} that use other, independent portions of the
atmospheric neutrino spectrum.  But these results themselves are at odds
with the Kamioka multi GeV results\cite{kmulti}.  While the oscillation
hypothesis can not be completely ruled out it is reasonable to seek alternative
explanations for the observations.

These experiments are sensitive to extremely low energy densities, which
have never been probed before.  Since there is no way to tell whether
the observed signal is actually  attributable to neutrinos, nor if they
are neutrinos if they are produced in the atmosphere it is prudent to consider
other possible alternatives.  In particular a flux of any particle that
interacts in a way that does not produce the distinctive energetic muon
in the final state would contribute to an increase in the relative rates of
observed ``electron'' like events to muon neutrino induced events.

This note explores a number of constraints that can be placed on sources
of non muon type interactions in underground detectors.

\section{Cross Section}
These events are observed in deep underground detectors.  This means that
whatever is producing them must have penetrated the earth (or have been
present in the detector beforehand).  There is no evidence for any directional
dependence to the event rates so that the upward going flux must be comparable
to that from all other directions.  The interaction length must be comparable
to the earth's diameter, or greater.
\[
L=\frac{1}{\sigma \rho N_{A}}
\]
The interaction length depends on the cross section and density traversed.
From the known density of the earth we can get a bound that
$\sigma < 2.4 \times 10^{-34} cm^2$.  This is a conservative bound, in that we
have used the average density of the earth but the upward going particles
would have traversed the core which has considerably more mass.
This cross section limit is well above the cross section for 1 GeV neutrinos
which is about $0.7 \times 10^{-38} cm^2$.

One might argue that while the observed events are isotropic this does not
necessarily require an isotropic flux.  Rather a reaction yielding an isotropic
energy flow would do.  But the majority of the observed events are classified
as ``single prong'' implying that the observed energy and momentum are
comparable.  The momentum must be brought in with the interacting particle
and so its flux is most likely isotropic.  While this argument becomes more
reliable as the energy increases the absence of anisotropy at any energy
makes it the simplest interpretation.

\section{Excess Flux Limit}
If we attribute the anomalous observations to the presence of an excess of
a non muon producing type of event we can get a bound on this new flux
from the observed event rate and the cross section bound we have estimated.
The observed muon flux is about 60\% of its expected value relative to the
non muon component.  If the observed depression of the muon to electron
ratio is interpreted as an enhancement in the ``electron'', {\em i.e.}
non muonic component this enhancement must account for from 20\% to 30\% of
all of the
observed events.  The event rate in a 3.3 kiloton water detector is about 1
event per day\cite{rate}, which is about
$R=5.8 \times 10^{-39}$ events/second/Nucleon.
This yields a flux limit of
\[
F_{DM} > P R/ \sigma
\]
where P is the fraction of all events attributable to new physics and $F_{DM}$
is the flux of new (dark) matter.

Here any possible nuclear shadowing has been neglected.  It is assumed that all
of the target nucleons available for neutrino interactions are available for
this new interaction too.  While one might argue that the lack of a
significant observed atmospheric neutrino anomaly in iron detectors\cite{iron}
might imply some
shadowing in the heavier iron nucleus the upper bound we have found for the
cross section, in the earth, makes this unlikely.

The limit on flux obtained from these arguments is
$F_{DM} > P \times 2.4 \times 10^{-5}$ particles/cm$^{2}$/second.
With $P=0.25$ this is $F_{DM} > 6 \times 10^{-6}$ particles/cm$^{2}$/second.

Recall that the flux is
inversely proportional to the cross section.
These additional particles could be electron neutrinos that are
not of atmospheric origin.  Using an average neutrino
cross section of $\sigma_{\nu_{e}}=3.4 \times 10^{-39}$ cm$^{2}$,
$<E_{\nu}>$=500 MeV the excess flux would be 
$F_{\nu_{e}}=0.4$ neutrinos/cm$^{2}$/second.

Only a lower bound on the flux of new particles can be obtained from this
argument since if it is a new interaction of a new particle the cross
section is not known.

\section{Energy Density}
A continuous flux of new particles would indicate the presence of an
energy density.  The energy density can be estimated from
\[
\epsilon = F_{DM} <E> / v
\]
where $<E>$ is the mean energy and
$v$ is the velocity of the flux.  The mean energy can be estimated from the
energy deposition by the interaction.  But the visible energy found in the
detector is usually a fraction of the energy of the incident particle.
For $\nu_{e}$ interactions the visible energy is equal to the energy of the
electromagnetic shower produced but on average this is only 1/2 of the
neutrino energy.  The observed visible energy distribution of the excess
events seems to be flat below about 600 MeV\cite{manpdk}.  We will take
$<E> = \frac{300}{x}$ MeV, where x is the fraction of particle energy
found in the detector.  It is assumed that the velocity is of the order
the speed of light, that is the particles are relativistic.
This yields
$\epsilon > 2.4 \times 10^{-13} \frac{P}{x}$ MeV/cm$^{3}$.  This is about
$4 \times 10^{-19} \frac{P}{x}$ ergs/cm$^{3}$, which should be compared with
the cosmic matter density of one nucleon per cubic meter.
$\epsilon_{cosmic} = 1.5 \times 10^{-9}$ ergs/cm$^{3}$.

If the events are assumed to be $\nu_{e}$ induced so that $\sigma$, $x$ and $P$
are known one gets
$\epsilon_{\nu_{e}}=8.6 \times 10^{-9}$ MeV/cm$^{3}$ or
$\epsilon_{\nu_{e}}=1.4 \times 10^{-14}$ ergs/cm$^{3}$ which is
well below the cosmic baryon energy density.

The choice of $<E> = \frac{300}{x}$ MeV is conservative.  It is possible
that the anomaly does not extend to higher energies.  Even if the evidence
presented in reference\cite{kmulti} is correct the mean energy of the flux
will only be higher.  Higher mean energies for this ``dark matter'' would
lead to tighter bounds than those presented here.  It is possible that
if the new matter responsible for the
anomaly has a cross section that drops rapidly with energy there could be
considerably more of it present than as sampled by the observed effect.
Using a, possibly low, energy estimate based on observations makes
the limits obtained conservative.

\section{Cross Section Lower Bound}
Given a bound on the energy density of the universe we can get a lower limit
on the cross section if this energy density is manifesting itself via these
excess underground events.  The relationship between energy density and cross
section can be summarized by
$\epsilon = \frac{P R}{\sigma} \frac{<E>}{v}$
where R is the number of events observed per nucleon per second,
P is the fraction of events attributable to the new particle,
$\sigma$ is the interaction cross section, $<E>$ is the mean energy
of the interacting particles and $v$ is the velocity of these particles.

One can expect an upper bound on the energy density to be enough to close
the universe, $\epsilon_{closure}$.  Under these conditions one finds
$\sigma_{min} > \frac{P R}{\epsilon_{closure}} \frac{<E>}{v}$
With $\epsilon_{closure}=10^{-8}$ergs/cm$^{3}$\cite{weinb,kandt} this yields
$\sigma_{min} > 4.6 \times 10^{-45}$ cm$^{2}$/Nucleon for $P=0.25$,
$<E>=600$ MeV and $v=c$.

A crude flux bound can be obtained from some of these ideas.  It is dependent
only on the observed energy and the closure bound on energy density.
\[
F_{DM} < \epsilon_{closure} v / <E>
\]
To be conservative we take $v = c$ and $<E> >$300MeV, which yields
$F_{DM} < 6.2 \times 10^{5}$ particles/cm$^{2}$/second.

\section{Decays}
It is possible that the excess events observed as the anomaly come from
the decay of particles in the detector rather than interactions with an
ambient flux.  This hypothesis is attractive since the anomaly is
not confirmed by dense detectors but only by the relatively low density water
detectors.  Neutrino interactions (and proton decay) should depend of the
fiducial mass of the device.  But if one is observing the decay of an
ambient dark matter flux the rate should depend on the {\em volume} of  the
detector and not its mass.  The low density water detectors observe a
significantly large volume, by a factor of 3 to 4 relative to the mass
viewed.  So the anomalous fraction of decay events found in dense detectors
should be greatly suppressed relative to neutrino interactions.

The observed decay rate should be
$R_{u}=\frac{\rho_{N} V}{\tau}$ where V is the volume of the detector,
${\rho_{N}}$ is the number density of the decaying particle and $\tau$ is the
particle lifetime.  For water detectors the excess event rate per unit volume
$R_{u}/V = P \times 3.5 \times 10^{-15}$ events/second/cm$^{3}$.  $P$ is
the fraction of total events attributable to dark matter decay, about 25\%.

For a bound on this hypothesis we can rewrite it in terms of the energy
density $\epsilon$ and the particle lifetime $\tau$.
\[
\frac{\rho_{N}}{\tau} = \frac{\epsilon}{<E> \tau} = \frac{R_{u}}{V}
\]
\[
\frac{\epsilon}{\tau} = \frac{R_{u} <E>}{V}
\]
Here $<E>$ is the mean energy associated with each particle at the present
time.  Such particles might be very massive but we can bound $<E>$ by
the average energy observed in the detector for each decay $<E>$ $>$ 300 MeV.
This implies:
\[
(\frac{\epsilon}{\tau})_{Obs} > 8.4 \times 10^{-19} ergs/cm^{3}/sec.
\]

One expects an upper bound on $\epsilon$ to be $\epsilon_{closure}$ and a
lower bound on $\tau$ to be comparable to the age of the universe.  This yields
\[
\frac{\epsilon}{\tau} < \frac{\epsilon_{closure}}{\tau_{universe}}
= 3.2 \times 10 ^{-26} ergs/cm^{3}/sec
\]

Comparing this with the observational result above we can conclude that the
hypothesis of the decay of ambient dark matter can be ruled out.  This bound
could be circumvented if there is a reason why the dark matter should cluster
at well above the cosmic density limit in the vicinity of the detector.

\section{Conclusions}
Since the origin of the ``atmospheric neutrino anomaly'' is still
uncertain we have attempted to place a number of bounds on the possible
source.  While the bounds include the conventional explanation of muon
neutrino oscillations, our more general approach gives a range of
alternatives and may provide motivation for additional theoretical
and experimental work on the subject.

The dark matter we have set limits on refers to an ambient, weakly interacting
form of matter.  It is clear from the work in section IV on energy density
that at the low density limits permitted by this work, there would be
negligible gravitational effects on galactic dynamics.  On the other hand
we have used closure of the universe as other analyses have\cite{cowsik}
to get an upper bound on the energy density where gravitational effects would
certainly be noticed.

The decay of a primordial homogeneous component can be ruled out.  The absence
of any apparent point sources, in terrestrial or celestial coordinates, in
the data implies either a diffuse local source, or a cosmological one.  It is
difficult to understand how natural processes could populate the several
hundred MeV energy region with electron neutrinos or other penetrating
particles.

\section*{Acknowledgements}
I would like to thank John Poirier for comments on the manuscript.

\end{document}